\documentclass{article}
\usepackage{times}
\usepackage{graphicx} 
\usepackage{subfigure} 

\usepackage{natbib}

\usepackage{algorithm}
\usepackage{algpseudocode}
\usepackage{graphicx}
\usepackage{amsmath,amssymb,amsthm,bm,enumerate}
\usepackage{latexsym,color,verbatim,minipage-marginpar,caption,multirow}
\usepackage{multicol}
\usepackage{array}
\usepackage{soul}

\newcommand{\eat}[1]{}

\theoremstyle{definition}

\DeclareFontFamily{U}{mathx}{\hyphenchar\font45}
\DeclareFontShape{U}{mathx}{m}{n}{
      <5> <6> <7> <8> <9> <10>
      <10.95> <12> <14.4> <17.28> <20.74> <24.88>
      mathx10
      }{}

\long\def\comment#1{}

\algnewcommand{\IfThen}[2]{
	\State \algorithmicif\ #1\ \algorithmicthen\ #2\ }

\author{Daniel Ting}
\title{Simple, Optimal Algorithms for Random Sampling Without Replacement}

\begin{document} 
\maketitle	
	\begin{abstract}
	Consider the fundamental problem of drawing a simple random sample (SRS) of size $k$ without replacement from $[n] := \{1, \ldots, n\}$.	
Although a number of classical algorithms exist for this problem, we construct algorithms that are even simpler, easier to implement, and have optimal space and time complexity. 
\end{abstract}
	
	\section{Introduction}
Sampling without replacement is a fundamental operation. We present several improvements to existing algorithms which make them both simpler to implement and have optimal time and space complexity. 
In contrast, existing classical algorithms used in practice, such as
 sampling with hash based membership checking and classical swapping, 
are typically suboptimal for some inputs.  They also covers several scenarios of particular interest, including sampling sequentially when data can only be accessed in order and distributed sampling. Table \ref{tbl:algos} gives a list of new algorithms and the scenarios in which they apply.

\begin{table}
	\begin{tabular}{r|llccl}
		Algorithm & Sample order & Output & n & k & Scenarios \\ \hline
		Fisher & Random & Sequence & Given & - & - \\
		Membership checking & Random & Sequence & Given & - & - \\
		\bf{Sparse Fisher-Yates} & Random & Sequence & Given & - & Unknown k\\
		\bf{Pre-initialized FY} & Random & Sequence & Given & - & Given sorted array\\
		\bf{Beta-Binomial} & Sorted & Sequence & Given & Given & In order, Distributed\\
		Floyd's & Biased & Array & Given & Given & - \\
		Reservoir & Random & Array & - & Given & Unknown n \\
		HyperGeometric & Blockwise & Blockwise & Per Block & Given & Distributed 
	\end{tabular}
	\caption{Type of output and inputs needed for the basic version of each algorithm, and special scenarios each is well suited for. 
	}
\label{tbl:algos}
\end{table}

We show how these methods, as well as existing ones, are all related. Each method is a natural consequence of a method for generating permutations 
and a representation of the permutation. 

\section{Fisher-Yates sampling}
Consider the classical swapping, or Fisher-Yates, method for generating a simple random sample given below:
\begin{algorithm}[H]
\caption{Fisher Yates Sampler(n,k)}
	\begin{algorithmic}
		\State {$x \gets 1,\ldots, n$}
		\For{$i = 0 \to k-1$}
		\State $r \gets \textit{Uniform}([n-i])$ 
   	     \State $Swap(x[n-i], x[r])$
		\EndFor
		\State \Return $x[(n-k+1):n]$
	\end{algorithmic} 
\end{algorithm}

In the classical swapping procedure, the last $i$ elements in the array form a random sample without replacement
after the $i^{th}$ iteration. The first $n-i$ elements consist of all items that have not been sampled. By performing a swap, the algorithm moves an item from the remaining items that havae not been sampled and adds it to the sampled items.
Thus, each iteration exactly mimics the process of sequentially sampling without replacement.
\begin{enumerate}
    \item Sample index $r$ from remaining items
    \item Remove $x[r]$ from the remaining items, and move it to the sampled items
\end{enumerate}

This algorithm  takes both $O(n)$ time and space since it initializes an array of length $n$.
This initialization step is wasteful when  $k \ll n$ is small. In this case, at most $2k$ locations in the array can be affected by swaps. A majority of locations in the array contain no meaningful information and simply store $x[i] = i$.

Our first improvement is given in algorithm \ref{alg:cyclesample} which replaces this dense representation of the array with a sparse representation using a hash table.
Only entries that have been swapped are stored, in other words, those with $x[i] \neq i$, and entries in the hash table can be deleted once it is no longer possible to select them. Since only the representation of the permuted array changes, given the same sequence of random bits, the algorithm produces exactly the same random sample as the classical swapping procedure.


	\begin{algorithm}[H]
	\caption{Sparse Fisher-Yates Sampler(n,k)}
	\label{alg:cyclesample}
	\begin{algorithmic}
		\State $H \gets HashTable()$
		\State $x \gets array(k)$
		\For{$i = 0 \to k-1$}
		\State $r \gets \textit{Uniform}([n-i])$ 
		\State $x[i] \gets H.get(r, \mathit{default}=r)$
		\State $H[r] \gets H.get(n-i, \mathit{default}=n-i)$		 
		\IfThen{$r = n-i$}{Delete $H[n-i]$}	 \qquad (Optional)
		\EndFor
		\State \Return $x$
	\end{algorithmic}
\end{algorithm}

Clearly the method takes $O(k)$ time and space and thus has optimal time and space complexity for drawing a random sample of size $k$ from $n$ items. Furthermore, it takes exactly k random draws. It is also particularly useful for a variation of the simple random sampling problem where the number of items $n$ is specified in advance, but the number of needed samples $k$ is not known.
Since the procedure mimics the natural process of sampling without replacement, 
one can implement the algorithm as an iterator that is initialized using just $n$
and which generates a sequence of draws without replacement. This can be useful, for example, in generating mean estimates with that are guaranteed to satisfy a desired accuracy. In this case, it is not known a priori how many samples $k$ are needed to achieve the desired accuracy, but using the sparse swapping method, one can repeatedly sample from the data until the estimated variance of the mean estimate is less than some desired maximum error. 


Algorithm \ref{alg:cyclesample} also bears a strong resemblance to one of the other main classical sampling methods, sampling with hash-based membership checking  \citep{ernvall1982algorithm}.
This method samples items with replacement 
and discards duplicates until $k$ distinct items are selected. Its pseudocode is given in algorithm \ref{alg:hashsample}. 


	\begin{algorithm}[H]
	\caption{Membership Checking Sampler(n,k)}
	\label{alg:hashsample}
	\begin{algorithmic}
		\State $H \gets HashTable()$
		\State $x \gets array(k)$
		\For{$i = 0 \to k-1$}
		\State {\textbf{repeat} \,$r \gets \textit{Uniform}([n])$\,\textbf{until} \,$r \not \in keys(H)$}
		\State $x[i] = r$
		\State $H[r] = 1$	
		\EndFor
		\State \Return $x$
	\end{algorithmic}
\end{algorithm}

The primary difference is that while membership checking must resample if an index has already been chosen,
the sparse swapping method stores a replacement index for each previously chosen index. Rather than resampling, the unique replacement index replaces an index that is sampled more than once.

\subsection{Time and Space Complexity}
Classical swapping takes $O(n)$ time and space due to the need to initialize an array of length $n$.
After sampling $t$ items, the probability hash based membership checking chooses a repeat item is $t/n$. The number of times it resamples to draw a new item is $Geometric(1-t/n)$ which has expect mean $n/(n-t)$. The expected number of times it performs a draw with replacement is thus $\sum_{t=0}^{k-1} n/(n-t) = n (H_n - H_{n-k}) \approx n \log(n/(n-k)) =
-n \log (1-k/n)$. Thus, it has time complexity $O(-n \log (1-k/n))$. Trivially, the space complexity is $O(k)$. 
Sparse swapping takes optimal $O(k)$ time and space and draws precisely $k$ uniform random variates.

When implemented as an iterator, we require a more refined space analysis accounting only 
for the size of bookkeeping data structures and not 
the output array of size $O(k)$.
It is easy to see that the number of hash entries after iteration $i$ is equal to the number of selected indices that are $< n-i$. 
The probability of never selecting index 1 is  $\frac{n-1}{n} \frac{n-2}{n-1} \ldots {n-i}{n-1+1} = \frac{n-i}{n}$. Since there are $n-i$ possible indices, the expected number of selected indices is
$\frac{(n-i)i}{n} = n q(1-q)$ where $q = i/n$. Thus, when $i \ll n$, the number of hash entries is $\approx i$ as expected. By maximizing over $q$, the maximum expected number of hash entries is $n/4$. This maximum size is attained when half the items are sampled, $i=n/2$. 

Table \ref{tbl:complexity} summarizes the expected time and space complexity of our sampling algorithms and other sampling algorithms used in practice. For space complexity, we show the space required to generate, but not store, $k$ draws without replace. Since random number generation is often the most expensive operation in sampling procedures, we also provide the expected number of random variates needed.

\begin{table}
	\begin{tabular}{r|ccc}
		Algorithm & Time  & Space  & Random draws \\ \hline
		Classical Swapping & $O(n)$ & $O(n)$ & $k$ \\
		Membership checking & $O\left(n \log\left(\frac{n}{n-k}\right)\right)$ & $O(k)$ & $n \log\left(\frac{n}{n-k}\right) + O(1)$  \\
		HyperGeometric & $O(k \log k)$ & $O(\log k)$ & $O(k \log k)$  \\
		Reservoir Sampling & $O(n)$ & $O(k)$ & $n$ \\
		Reservoir Sampling w/ Skipping & $O(k \log n)$ & $O(k)$ & $k \log n + O(1)$ \\
		Selection Sampling & $O(n)$ & $O(1)$ & $n$ \\ 
		\hline	
		Sparse Swapping & $O(k)$ & $O(k (1-k/n))$ & $k$  \\
	 	Pre-initialized FY w\slash undo & $O(k)$ & $O(1)$ & $k$ \\ 
		Beta-Binomial& $O(k)$ & $O(1)$ & $k$ \\		
	\end{tabular}
	\caption{Table of algorithms and their complexity}
	\label{tbl:complexity}
\end{table}

\subsection{Product of transpositions and Pre-initialized Fisher-Yates sampling}
Since the main cost of the classical swapping method is the initialization, we consider a small variation that ensures the array is left in the same state as it was in the beginning.
Given an input array, we sample from the array using the classical swapping method but then undo the swaps.

To see how this can easily be done, we first consider a mathematical representation of the  Fisher-Yates shuffling algorithm as a product of transpositions. 
The swapping method and the Fisher-Yates shuffle for generating a permutation can be represented mathematically as a product of transpositions where each transposition swaps items in an array. Specifically, a permutation $\pi$ can be uniquely represented by a product
\begin{align}
	\pi = (1\; r_i) (2\; r_2) \cdots (n\; r_n)
\end{align}
where $r_i \leq i$. Thus, a uniform random permutation can be drawn by taking independent draws of $r_i \sim Uniform([i])$.
The Fisher-Yates shuffle is the left action of a random permutation in this representation on an array. The action of a transposition $(i\; r_i)$ on an array simply swaps the items at positions $i$ and $r_i$.
The left action applies the transpositions in this representation from right to left. 
Interestingly, reservoir sampling is natural consequence of applying the \emph{right} action of this representation and maintaining only the first $k$ elements of the array.

The classical swapping method truncates the permutation and only applies $k$ transpositions. This takes advantage of a stability property of this representation. The $n-k$ leftmost transpositions leave positions $n-k+1$ to $n$ untouched. Since classical swapping only uses these $k$ positions at the end for the sample, it does not need to apply the $n-k$ remaining transpositions that do not affect the end of the array. 

Thus, undoing the permutation $((n-k+1)\;\, r_{n-k+1}) \cdots (n\; r_n)$ applied by classical swapping simply requires applying the transpositions in reverse order. One can simply store the values $r_{n-k+1}, \ldots, r_n$ and apply $(n\; r_n) \cdots ((n-k+1)\;\, r_{n-k+1})$.

\begin{algorithm}[H]
\caption{Pre-initialized Fisher-Yates with undo(x, n, k)}
\begin{algorithmic}
    \State $U \gets array(k)$
    \State $sample \gets array(k)$
	\For{$i = 1 \to k$}
		\State $r \gets \textit{Uniform}([n-i])$ 
   	     \State $Swap(x[n-i], x[r])$
   	     \State $U[i-1] \gets r$
   	     \State $sample[i-1] \gets x[n-i]$
		\EndFor
		\For{$i = k \to 1$}
		\State $r \gets U[i-1]$
		\State $Swap(x[n-i], x[r])$
		\EndFor
		\State \Return $sample$
\end{algorithmic}
\end{algorithm}

\section{In order sampling}
The methods above sample items in random order. In cases where data is accessed linearly, it is useful to return indices for sampled items in sorted order. This is also referred to as sequential sampling. 

The basic in-order sampling method, also known as selection sampling, iterates through indices and for each index, randomly selects it with probability $k_{\mathit{left}}/n_{\mathit{left}}$ for the sample where $k_{\mathit{left}}$ 
is the number of remaining samples to be drawn and 
and $ n_{\mathit{left}}$ is the number of remaining items. It draw $n$ random variates and takes $O(n)$ time. 

\begin{algorithm}[H]
\caption{Selection Sampler(n,k)}
	\begin{algorithmic}
				\State $x \gets List()$				
				\For{$i \gets 1 \to n$} 
				\If{$Bernoulli(k/(n-i+1)) = 1$}
					\State $x. \mathit{append}(i)$
					\State $k \gets k -1$
				\EndIf 
				\State \Return x
	\EndFor
	\end{algorithmic}
\end{algorithm}

This can be made more efficient by drawing how many items to skip rather than iterating through them.  \cite{devroye1981} and \cite{vitter1987efficient} provide efficient, in order sampling algorithms with skipping. However, they are infrequently used in practice as \cite{NonUniformGeneration} states that they are non-trivial to implement . Table \ref{tbl:implementations2} show the algorithms used by a number of popular libraries and languages. Only Julia implements an in order sampling procedure with skipping, namely Algorithm D in \cite{vitter1984faster}.

\begin{table}
	\begin{tabular}{ll|c|l}
	Software & Version & Time & Algorithms \\ \hline
	\emph{numpy}  & 1.20.1 & $O(n)$ & {Full Fisher-Yates shuffle} \\
	\emph{gsl} & 2.6 & $O(n)$ & {Basic sequential} \\
\emph{Commons (Java)} & 3.6.1 & $O(n)$ & {Classical swapping} \\
	\emph{C++14} & libstdc++ 6.0.29 & $O(n)$ & {Reservoir, Selection} \\
	\emph{Spark} & 3.1 & $O(n)$ & Full Shuffle, Reservoir, Oversampling \\ 
	R & 4.04 & $O(k)$ & Classical swapping  + Membership checking  \\
	Rust & 1.50.0 & $O(k)$ & Classical swapping + Membership checking \\
	Julia & 1.5.4 & $O(k)$ & Swapping + Membership checking + Reservoir + \bf{Vitter Algo. D} 
\end{tabular}


\caption{Common software and their sampling algorithms. Software where only $O(n)$ algorithms are available are \emph{italicized}. A $+$ denotes a polyalgorithm that chooses the base sampling implementation based on the inputs. Only Vitter's algorithm D is an  $O(k)$ base algorithm.}
\label{tbl:implementations2}
\end{table}

Our contribution is to show that an in order sampler with skipping can be easily implemented in a few lines of code if a $Beta-Binomial$ or $Binomial$ distributed random number generator is available. From the selection sampling algorithm, it is easy to see that the probability that the first sampled item $X_1$ is equal to $x$ is
\begin{align}
    P(X_1 = x) &= \frac{k}{n-x+1}\prod_{i=1}^{x-1} \frac{n-i+1-k}{n-i+1} \\
    &= k \frac{(n-k)!}{(n-k-x-1)!}\frac{(n-x)!}{n!} = \frac{{n-x \choose k-1}}{{n \choose k}}.
\end{align}
Previous approaches have treated this distribution as a special distribution and developed 
specialized algorithms to sample from the distribution. We show that this is a standard $\mathit{Beta-Binomial}(1,k,n)$ distribution. Thus, if a $Beta-Binomial$ sampler is readily available, implementing the in order sampler is trivial.
 Furthermore, the $Beta$ random variates are particularly simple to sample from, so implementation is simple as long as a $Binomial$ sampler is available.

While identification of the distribution as a $Beta-Binomial(1, k, n)$ can be verified algebraically, we provide a combinatorial proof that allows us to characterize the locations of all the sampled items jointly.
Consider the following process. Start with a randomly permuted array and select the first $k$ items. Now sort the array. The locations of the selected items are a simple random sample without replacement of $[n]$. It is easy to see this by starting with a random permutation of $1, \ldots, n$. The locations are equal to the selected items themselves.

Suppose instead that we start with $n$ i.i.d. $Uniform(0,1)$ random variables $U_1, \ldots, U_n$. 
Denote the order statistics formed from sorting the $U_i$ by $U_{(1)} < \ldots < U_{(n)}$. 
Since the locations of $U_1, \ldots, U_k$ in this sorted array generate a simple random sample, the locations of their sorted values $\tilde{U}_{(1)}, \ldots, \tilde{U}_{(k)}$ yield an in order sample. 

To derive the distribution of the locations, note that 
the location $X_i$ of $\tilde{U}_{(i)}$ is simply the total number of random variates $U_j \leq \tilde{U}_{(i)}$.
Trivially, there are $i$ such values among the selected $U_1, \ldots, U_k$. The number of values among the remaining random variates $U_{k+1}, \ldots, U_n$ is
\begin{align}
    X_i - i | \tilde{U}_{i} &\sim Binomial(n-k, \tilde{U}_{i}).
\end{align}
The value of $\tilde{U}_1$ is the minimum of $k$ $Uniform(0,1)$ random variables. One can either simply note that this is $Beta(1, k)$ distributed, or derive from first principles
that $P(\tilde{U}_1 \geq z) = (1-z)^k$, which has the desired distribution. 
Thus, $X_1 \sim Beta-Binomial(1, k, n-k) + 1$. 
This can be generalized to find the joint distribution for the number of random variates 
\begin{align}
    \tilde{U}_{(1)}, \ldots, \tilde{U}_{(k)} &\sim Dirichlet(1, \ldots, 1) \\
    (\Delta_1,\, \Delta_2 ,\, \ldots, \,\Delta_k,\, \Delta_{k+1}) | \tilde{U} &\sim Multinomial(n-k, \tilde{U}_{(1)}, \tilde{U}_{(2)}, \ldots, \tilde{U}_{(k)}, 1- \tilde{U}_{(k)}) \nonumber
\end{align}
where $\Delta_i = X_i - X_{i-1} - 1$ with $X_0 = 1$ and $X_{k+1} = n$.

\begin{algorithm}
\caption{Beta-Binomial In Order Sampler}
\begin{algorithmic}
    \Function{BetaBinomialSampler}{n,k}
        \State $x \gets array(k)$
        \For{$i \gets 1 \to k$}
            \State $x[i] \gets Beta-Binomial(1,k,n-k) + 1$
            \State $n \gets n-x[i]$
            \State $k \gets k-1$
        \EndFor
        \State \Return $x$
    \EndFunction
    \Function{Beta-Binomial}{$\alpha=1, \beta, n$}
        \State $U \gets \mathit{Uniform}(0,1)$
        \State $p \gets 1-U^{1/\beta}$
        \State \Return $Binomial(n, p)$
    \EndFunction
\end{algorithmic}
\end{algorithm}

\section{Distributed sampling}
When sampling from large databases, data may not only be on disk, but also distributed across machines. In this distributed setting, \cite{chickering2007distributed} introduced a divide-and-conquer approach using HyperGeometric random variables to decide how many samples should be drawn from each machine. \cite{Sanders2018EfficientPR} rediscovered this algorithm and use it in additional scenarios. However, implementations of hypergeometric random number generators are relatively rare.
Rather than fix the size of the sample but using simple Bernoulli or geometric random variables, \cite{Meng2013} proposed an oversampling approach that ensures enough samples are chosen with high probability.

The Beta-Binomial in order sampler can be adapted to the distributed case. While the in order sampler samples the position of the first item in the sample, the distribution for the position of any item in the sample can be found just as easily. The $j^{th}$ sampled item's position is distributed $Beta-Binomial(j, k-j+1, n-k)+j$.

This can be used, for example, to implement a simple hypergeometric random number generator via a search procedure. A $Hypergeometric(v,n,k)$ can be interpreted as finding how many items in a sample of size k occur in the first $v$ positions. One can guess which sampled item occurs close to the $v^{th}$ position and draw a $Beta-Binomial$ to find its exact position. If the guess is $j$ and its position is $\ell$, then $j$ is a lower bound on the hypergeometric random variable if $\ell \geq v$ and an upper bound otherwise. Successive draws tighten the bound and can find the exact value.

\subsection{Merging samples}
A novel application of the method is to merge samples obtained from multiple nodes in a network or from multiple data sets.
Consider data sets $\mathcal{D}_1, \mathcal{D}_2$ of size $n_1, n_2$ respectively and
simple random samples $\mathcal{J}_1,  \mathcal{J}_2$ with size $k_1,k_2$.

To obtain a simple random sample from $\mathcal{D}_1 \cup \mathcal{D}_2$
assign an auxiliary $Uniform(0,1)$ random variable to each of the $n_1 + n_2$ items and take the items with smallest random values.
Denote the sorted uniform random values for $\mathcal{D}_c$ by $U_{(1)}^c,  U_{(2)}^c, \ldots, U_{(n_c)}^c$.
Since only items with random value less than $T_c := U_{(k_c+1)}^c$ are observed in the sample $\mathcal{J}_c$, 
a random sample of the union can only contain items with auxiliary random value less than $T' := \min \{ T_1, T_2 \}$.
Generate a merged random sample by computing the number of auxiliary variables with value less than $T'$ for each dataset. 
Denote this number by $\kappa_c$ for data set $\mathcal{D}_c$. The merged random sample consists of the union random samples of size $\kappa_c$ 
from $\mathcal{J}_c$.
These can be generated by taking
\begin{align}
T^1 &\sim Beta(k_1+1, n_1 - k_1) \\
\kappa_1 &\sim Binomial(k_1, \min\{1, T_1 / T_2 \})
\end{align}
and similarly generating $\kappa_2$.
This produces a merged simple random sample of maximum possible size $\kappa_1 + \kappa_2$ which can be further downsampled if necessary. This is summarized in algorithm \ref{alg:distributed}

\begin{algorithm}
\caption{Merging Distributed Samples}
\label{alg:distributed}
	\begin{algorithmic}
		\Function{MergeSamples}{$\mathcal{J}_1, \mathcal{J}_2, n_1, n_2$}
		\State $k_1, k_2 \gets |\mathcal{J}_1|, |\mathcal{J}_2|$
		\State $T_1 \sim Beta(k_1+1, n_1 - k_1)$
		\State $T_2 \sim Beta(k_2+1, n_2 - k_2)$
		\State $\kappa_1 \sim Binomial(k_1, \min\{1, T_1 / T_2 \})$
		\State $\kappa_2 \sim Binomial(k_2, \min\{1, T_2 / T_1 \})$	
		\State $\mathcal{J}_1' \gets Sample(\mathcal{J}_1, \kappa_1)$
		\State $\mathcal{J}_2' \gets Sample(\mathcal{J}_2, \kappa_2)$
		\State \Return $\mathcal{J}_1' \cup \mathcal{J}_2'$
		\EndFunction
	\end{algorithmic}
\end{algorithm}

\bibliographystyle{abbrvnat}
\bibliography{ling2}

\begin{thebibliography}{8}
\providecommand{\natexlab}[1]{#1}
\providecommand{\url}[1]{\texttt{#1}}
\expandafter\ifx\csname urlstyle\endcsname\relax
  \providecommand{\doi}[1]{doi: #1}\else
  \providecommand{\doi}{doi: \begingroup \urlstyle{rm}\Url}\fi

\bibitem[Chickering et~al.(2007)Chickering, Roy, and
  Meek]{chickering2007distributed}
D.~M. Chickering, A.~K. Roy, and C.~A. Meek.
\newblock Distributed reservoir sampling for web applications, Dec.~11 2007.
\newblock US Patent 7,308,447.

\bibitem[Devroye(1986)]{NonUniformGeneration}
L.~Devroye.
\newblock \emph{Non-Uniform Random Variate Generation}.
\newblock Springer-Verlag, New York, 1986.

\bibitem[Devroye and Yuan(1986)]{devroye1981}
L.~Devroye and C.~Yuan.
\newblock {Inversion with correction for the computer generation of discrete
  random variables}.
\newblock Technical report, McGill Univerity, 1986.

\bibitem[Ernvall and Nevalainen(1982)]{ernvall1982algorithm}
J.~Ernvall and O.~Nevalainen.
\newblock An algorithm for unbiased random sampling.
\newblock \emph{The Computer Journal}, 25\penalty0 (1):\penalty0 45--47, 1982.

\bibitem[Meng(2013)]{Meng2013}
X.~Meng.
\newblock Scalable simple random sampling and stratified sampling.
\newblock In \emph{ICML}, 2013.

\bibitem[Sanders et~al.(2018)Sanders, Lamm, H{\"u}bschle-Schneider, Schrade,
  and Dachsbacher]{Sanders2018EfficientPR}
P.~Sanders, S.~Lamm, L.~H{\"u}bschle-Schneider, E.~Schrade, and C.~Dachsbacher.
\newblock Efficient parallel random sampling—vectorized, cache-efficient, and
  online.
\newblock \emph{ACM Transactions on Mathematical Software (TOMS)}, 44:\penalty0
  1 -- 14, 2018.

\bibitem[Vitter(1984)]{vitter1984faster}
J.~S. Vitter.
\newblock Faster methods for random sampling.
\newblock \emph{Communications of the ACM}, 27\penalty0 (7):\penalty0 703--718,
  1984.

\bibitem[Vitter(1987)]{vitter1987efficient}
J.~S. Vitter.
\newblock An efficient algorithm for sequential random sampling.
\newblock \emph{ACM transactions on mathematical software (TOMS)}, 13\penalty0
  (1):\penalty0 58--67, 1987.

\end{thebibliography}

\end{document}